\documentclass[prd,aps,a4paper,eqsecnum,floatfix,twocolumn]{revtex4}  

\usepackage{graphicx}
\usepackage{amsmath,amsfonts,amssymb}
\usepackage{slashed}

\newcommand{\beq}{\begin{equation}}
\newcommand{\eeq}{\end{equation}}
\newcommand{\bea}{\begin{eqnarray}}
\newcommand{\eea}{\end{eqnarray}}

\begin{document}

\title{Kerr spacetime and scalar wave equation: Exact resummation of the renormalized angular momentum in the eikonal limit
}

\author{Donato Bini$^{1}$,  Giorgio Di Russo$^{2}$, Andrea Geralico$^{1}$} 
  \affiliation{
$^1$Istituto per le Applicazioni del Calcolo ``M. Picone,''\\ CNR, I-00185 Rome, Italy\\
$^2$School of Fundamental Physics and Mathematical Sciences, Hangzhou Institute for Advanced Study, UCAS, Hangzhou 310024, China\\
}

\date{\today}

\begin{abstract}
We show that the null geodesic radial action for unbound orbits in the Kerr spacetime, and consequently the scattering angle, can be resummed in terms of hypergeometric functions, extending previous results [M.~M.~Ivanov, et al. arXiv:2504.07862]. 
We provide explicit expressions as series expansions in powers of the Kerr rotational parameter up the fourth order included. 
We finally use the Mano-Suzuki-Takasugi formalism to prove the relation between the renormalized angular momentum and the radial action highlighted in previous works. 
\end{abstract}

\maketitle

\section{Introduction}
\label{Intro}

Recently, Ref. \cite{Ivanov:2025ozg} has suggested that null particles radial action in the Kerr spacetime (at linear order in the rotational parameter $a$) can be related to the \lq\lq renormalized angular momentum" (RAM) appearing in the Mano-Suzuki-Takasugi (MST) formalism \cite{Mano:1996mf} for the analytical treatment of gravitational perturbations within the general Teukolsky Master Equation (TME) approach \cite{Press:1973zz}. 
See also Refs. \cite{Parnachev:2020zbr,Akpinar:2025huz}, as well as Ref. \cite{Nasipak:2024icb}
where  the monodromy data of
the radial TME have been related to the RAM (also providing a monodromy-based numerical scheme for calculating it, especially useful for computations in different regions of the parameter space). 
Successively, Ref. \cite{Bini:2025ltr} has  generalized these results to the case of Topological Star (TS) spacetime, clarifying the relation between Wentzel-Kramers-Brillouin (WKB)-type solutions of the radial equation (for a massless scalar field) and the RAM (see Refs. \cite{Bianchi:2025aei,Bini:2025qyn,DiRusso:2025lip,Bianchi:2024rod,Bianchi:2024vmi,Bena:2024hoh,Cipriani:2024ygw,DiRusso:2024hmd,Bianchi:2023sfs}
for recent applications to TS, including generalization of MST formalism to this case as well as the use of the companion quantum Seiberg-Witten formalism).

In the  present work we extend  the results of Ref. \cite{Ivanov:2025ozg} to higher orders in $a$, finding explicit  resummation expressions (in terms of hypergeometric functions) up to the fourth order in $a$, also confirming the relation between tha radial action and the  RAM of the  MST  formalism in the eikonal limit.
As soon as the expansion order in the rotational parameter of the black hole increases, these relations become more and more involved, and are written here as a (nontrivial) sum of hypergeometric functions (possibly to be further reduced by using the contiguity properties of the hypergeometric functions, a property that is still not well implemented in various algebraic manipulation software available today). 

We see then a flowing of information from (background) null geodesic radial action to (perturbation) RAM, in the sense that knowing the analytic representation of the radial action one gets nontrivial information about the MST  RAM (whose resummation was never attempted in the past years) and viceversa.

The paper is organized as follows: Sec. II is a short reminder of test particle motion in Kerr spacetime, within the Hamilton-Jacobi formalism and with special attention to equatorial null geodesics. Sec. III  introduces the radial action for null geodesics, while  Sec. IV discusses scalar wave motion, within the MST formalism and in the eikonal limit (postponing long expressions/technical details in Appendix \ref{rec_rel}). Conclusions are then summarized in Sec. V.

We use here the mostly positive convention for the metric signature and follow the notation of Ref. \cite{Misner:1973prb} (for both Kerr metric in Boyer-Lindquist coordinates and null geodesics).

\section{Test particle motion in a Kerr spacetime}
\label{kerrgeo}

The Kerr metric in standard  Boyer-Lindquist coordinates reads
\bea
ds^2&=&-\frac{\Delta}{\rho^2}\left(dt-as^2 d\phi\right)^2+\frac{s^2}{\rho^2}[a dt-(r^2+a^2)d\phi]^2\nonumber\\
&+&\rho^2\left(\frac{dr^2}{\Delta}+d\theta^2 \right) \,,
\eea
with 
\beq
\Delta=r^2-2M r+a^2\,,\quad \rho^2=r^2+a^2c^2\,,
\eeq
where $M$ and $a=J_{\rm bh}/M$ are the black hole mass and angular momentum (per unit mass), respectively, and the compact notation $[c,s]=[\cos\theta,\sin\theta]$ has been introduced.

Geodesic motion is conveniently studied by using the Hamilton-Jacobi formalism, which we briefly recall below.
Starting from the Lagrangian
\beq
{\mathcal L}(x^\mu, \dot x^\mu)=\frac12 g_{\mu\nu}(x^\alpha)\dot x^\mu \dot x^\nu\,,
\eeq
(a dot meaning derivatives with respect to an affine parameter $\lambda$) with canonically conjugate momenta 
\beq
p_\mu=\frac{\partial \mathcal{L}}{\partial \dot{x}^\mu}=g_{\mu\nu}\dot{x}^\nu\,,
\eeq
one can form the associated Hamiltonian
\bea
{\mathcal H}(x^\mu, p_\nu)&=& p_\mu \dot{x}^\mu(p_\nu)-{\mathcal L}(x^\mu, \dot x^\mu(p_\nu))\nonumber\\
&=&\frac12g_{\mu\nu}p^\mu p^\nu\,,
\eea
which is a functional of the coordinates $x^\mu(\lambda)$ and the conjugate momenta $p_\mu(\lambda)$.

The Hamilton-Jacobi approach consists in introducing a function $S(x^\mu,\lambda)$ such that $\partial_\mu S=p_\mu$, satisfying the equation
\bea
\label{HJ_eq}
0&=&\frac{dS}{d\lambda}+{\mathcal H}(x^\mu, \partial_\mu S)\nonumber\\
&=&\frac{dS}{d\lambda}+\frac12g^{\mu\nu}\partial_\mu S\partial_\nu S\,.
\eea
The solution of Eq. \eqref{HJ_eq} will depend, in general, on four integration constants.
Two constants of the motion, i.e., the conserved energy $E$ and angular momentum $L$, are associated with the Killing vectors $\partial_t$ and $\partial_\phi$ of the Kerr metric, respectively
\beq
p_t=-E\,,\qquad p_\phi=L\,.
\eeq
The third constant is the particle's rest mass $\mu$
entering the mass-shell condition 
\beq
\mathcal{H}=\frac12g^{\mu\nu}p_\mu p_\nu=-\frac12\mu^2\,.
\eeq
The fourth constant, known as Carter constant, is associated with the existence of a higher order symmetry of the Kerr spacetime generated by a second order Killing tensor field $K^{\mu\nu}$, 
\beq
{\mathcal K}=K^{\mu\nu}p_\mu p_\nu\,,   
\eeq
with
\beq
K^{\mu\nu}=-2\rho^2 l^{(\mu} n^{\nu)}+r^2 g^{\mu\nu}\,,
\eeq
where
\bea
l&=& \frac{(r^2+a^2)}{\Delta}\partial_t+\partial_r+\frac{a}{\Delta} \partial_\phi \,,\nonumber\\
n&=& \frac{1}{2\rho^2}[(r^2+a^2)\partial_t -\Delta \partial_r+a\partial_\phi]\,,
\eea
are null vectors aligned with the two repeated principal null directions of the Kerr spacetime (normalized so that $l\cdot n=1$) and $K_{(\mu\nu; \alpha)}=0$.
The Killing tensor $K^{\mu\nu}$ is directly related to the separability of the Hamilton-Jacobi equation \eqref{HJ_eq}.
Indeed, the latter equation can be  solved by separation of variables in the form
\beq
S=-\frac12\mu^2\lambda-Et+L\phi+S_r(r)+S_\theta(\theta)\,,
\eeq
with
\beq
S_r(r)=\int^r p_r dr\,, \qquad S_\theta(\theta)=\int^\theta p_\theta d\theta\,.
\eeq
We will limit our considerations below to the massless particle case $\mu=0$, implying 
\bea\label{separation}
p_\theta^2&=&Q-c^2 \left( \frac{L^2}{s^2}-a^2E^2\right)
\equiv\Theta(\theta)\,,\nonumber\\
p_r^2&=&\frac{(E(r^2+a^2)-a L)^2}{\Delta^2}-\frac{(L-aE)^2+Q}{\Delta}\nonumber\\
&\equiv& \frac{R(r)}{\Delta^2}\,,\nonumber\\
\eea
with ${\mathcal K}=Q+(L-aE)^2$.

\section{Radial action for unbound null geodesics in the equatorial plane}

We are interested here in the equatorial plane motion $\theta=\pi/2$, $p_\theta=0$, so that $Q=0$ implying ${\mathcal K}=(L-aE)^2$ and 
\beq
\label{radpot}
R(r)=(E(r^2+a^2)-a L)^2-\Delta(L-aE)^2\,.
\eeq
For notational reasons we will denote $L=J$ hereafter.

Upon introducing the dimensionless inverse radial distance $u={M}/{r}$ and impact parameter $\hat{b}={J}/{M E}$, the radial potential \eqref{radpot} $R(r)\vert_{r=M/u}\equiv R(u)$ becomes
\bea
R(u)
&=& E^2\left[  1 - u^2 (\hat{b} - {\hat a})\left(\hat{b}( 1-2u) + {\hat a}(1 + 2u) \right)\right]\,,\nonumber\\
\eea
with ${\hat a}={a}/{M}$.
The turning points are defined as the roots of the equation $R(u)=0$, which can be conveniently written as
\beq
\label{eqRv}
v^3+p v+q=0\,,
\eeq
in terms of the new variable 
\beq
v=u-\frac{\hat{b}+{\hat a}}{6(\hat{b}-{\hat a})}\,,
\eeq
where
\bea
p&=&-\frac{(\hat{b}+{\hat a})^2}{12(\hat{b}-{\hat a})^2}\,,\nonumber\\
q&=&-\frac{\hat{b}^3+3{\hat a} \hat{b}^2+3({\hat a}^2-18)\hat{b}+{\hat a}^3}{108(\hat{b}-{\hat a})^3}\,.
\eea
The roots of Eq. \eqref{eqRv} ordered so that $v_1>v_2>0>v_3$ are then given by
\bea
v_1&=&-\frac{(-1)^{2/3}p}{\mathcal{D}^{1/3}}-\frac{(-1)^{1/3}\mathcal{D}^{1/3}}{3}\,,\nonumber\\
v_2&=&-\frac{p}{\mathcal{D}^{1/3}}+\frac{\mathcal{D}^{1/3}}{3}\,,\nonumber\\
v_3&=&\frac{(-1)^{1/3}p}{\mathcal{D}^{1/3}}+\frac{(-1)^{2/3}\mathcal{D}^{1/3}}{3}\,,
\eea
with
\beq
\mathcal{D}=\frac32(\sqrt{12p^3+81q^2}-9q)\,,
\eeq
the corresponding roots $u_1$, $u_2$, $u_3$ having the same ordering when going back to the variable $u$.

For scattering orbits $0 < u \leq u_2$, with $u_2$ corresponding to the distance of closest approach $r_{\rm min} = M/u_2$. The condition $u_2=u_1$ gives the critical value of the impact parameter corresponding to capture by the black hole
\bea
\hat b_{\rm crit}&=&6\cos\left(\frac13{\rm arccos}(-\hat a)\right)-\hat a\nonumber\\
&=&3\sqrt{3}-2\hat a-\frac{\sqrt{3}}{6}\hat a^2-\frac{4}{27}\hat a^3-\frac{35\sqrt{3}}{648}\hat a^4+O(\hat a^5)\,,\nonumber\\
\eea
so that the Schwarzschild value $b_{\rm crit}^{\rm Schw}=3\sqrt{3}$ is reproduced for $\hat a=0$.
The PM approximation holds for values of the impact parameter much larger than the critical value, i.e., $\hat b\gg \hat b_{\rm crit}$.
For instance, the turning points for radial motion in a large $\hat{b}$ expansion result then
\bea
u_1&=&\frac{1}{2}+\frac{{\hat a} }{\hat{b}}+\frac{{\hat a}^2-2}{\hat{b}^2}+\frac{{\hat a} ^3+4 {\hat a} }{\hat{b}^3}+\frac{{\hat a} ^4-6 {\hat a} ^2-16}{\hat{b}^4}\nonumber\\
&+&O\left(\frac{1}{\hat{b}^5}\right)\,,  \nonumber\\
u_2&=&\frac{1}{\hat{b}}+\frac{1}{\hat{b}^2}+\frac{{\hat a}^2-4 {\hat a} +5}{2 \hat{b}^3}+\frac{3 {\hat a} ^2-10 {\hat a} +8}{\hat{b}^4}\nonumber\\
&+&O\left(\frac{1}{\hat{b}^5}\right)\,,  \nonumber\\
u_3&=&-\frac{1}{\hat{b}}+\frac{1}{\hat{b}^2}-\frac{{\hat a}^2+4 {\hat a}+5}{2\hat{b}^3}+\frac{3 {\hat a} ^2+10 {\hat a} +8}{\hat{b}^4}\nonumber\\
&+&O\left(\frac{1}{\hat{b}^5}\right)\,,
\eea
with $u_3(\hat b, \hat a)=u_2(-\hat b, -\hat a)$.

Let us define the action for unbound null geodesics as
\beq
I_r = \int_{r_{\rm min}}^{\infty} p_r dr\,.
\eeq
In terms of the $u$ variable it becomes
\beq
\label{Irnew}
I_r = \frac{J}{\hat b} \int_0^{u_2} du \mathcal{I}_r(u)
\,,
\eeq
where $\frac{J}{\hat b}=ME$, and
\beq
\mathcal{I}_r(u)=\frac{ \sqrt{  1 - u^2 (\hat{b} - {\hat a})\left[\hat{b}(1-2u) + {\hat a}(1 + 2u) \right]  } }{ u^2 \left(1 - 2u + {\hat a}^2 u^2 \right)}\,.
\eeq
The integration of \eqref{Irnew} can be formally represented in terms of special functions, e.g., Elliptic integrals, or equivalently, Lauricella hypergeometric functions \cite{Gonzo:2023goe}.
However, for the purposes of the present paper we will use below a PM expansion of the integrand (i.e., as a series expansion in the large impact parameter $\hat b$) for small values of the black hole spin parameter $\hat a$ (up to the order $O({\hat a}^4)$ included), showing that the series can be resummed at each order in $\hat a$, leading to closed expressions in terms of generalized hypergeometric functions.

In fact, introducing the new integration variable $z=\hat{b}u$, the radial action \eqref{Irnew} writes as
\beq
\label{Irnew2}
I_r=\hat{b}M E\int_0^{z_2} dz \mathcal{I}_r(z)\,,
\eeq
with $z_2=\hat bu_2=1+O(1/{\hat b})$, and
\beq
\mathcal{I}_r(z)=\frac{\sqrt{1-z^2+W(\hat a, \hat b; z)}}{z^2\left(1-\frac{2z}{\hat{b}}+\frac{{\hat a}^2 z^2}{\hat{b}^2}\right)}\,,
\eeq
with 
\beq
W(\hat a, \hat b; z)=\frac{2z^3}{\hat b}+\frac{z^2\hat a(-4z+\hat a)}{\hat b^2}+\frac{2z^3\hat a^2}{\hat b^3}\,.
\eeq
Both the upper limit of the integral and the integrand have a large-$\hat b$ expansion.
Expanding the integrand in powers of $\hat b$ will generate formally divergent integrals in both limits $z\to0$ and $z\to1$.
Following the prescriptions of Ref. \cite{Damour:2019lcq} one can get the correct result for the expanded integral by considering only the leading term in the PM expansion of upper limit (i.e., $z_2=1$), taking then the finite part of the divergent integrals at each order.
The radial action integral \eqref{Irnew2} thus becomes 
\beq
\label{Irfin}
I_r=\hat{b}M E \,{\rm Pf}\int_0^{1} dz \mathcal{I}_r(z)\,,
\eeq
where the integrand $\mathcal{I}_r(z)$ needs to be PM-expanded.
Notice that defining $z=u/u_2$ in Eq. \eqref{Irnew} also restricts the integral over $z$ between $0$ and $1$, leading to the same result.

We will proceed as follows.
Let us introduce the rescaled integrand
\beq
\mathcal{S}(z)=\frac{\sqrt{1+\frac{W(\hat a, \hat b; z)}{1-z^2}}}{1-\frac{2z}{\hat{b}}+\frac{{\hat a}^2 z^2}{\hat{b}^2}}\,,
\eeq
such that 
\beq
\label{conprefac}
\mathcal{I}_r(z)=\frac{\sqrt{1-z^2}}{z^2}\mathcal{S}(z)\,.
\eeq
Expanding then in powers of ${\hat a}$ (truncated here at $O(\hat a^4)$ included) gives
\beq
\mathcal{S}(z)=\sum_{k=0}^4\mathcal{S}_k(z)\hat a^k+O(\hat a^5)\,,
\eeq
where
\begin{widetext}
\bea
\label{Sn}
\mathcal{S}_0(z)&=& \mathcal{K}_{1/2,-1}(\hat b,z)\,,\nonumber\\
\mathcal{S}_1(z)&=& -\frac{2}{\hat{b}^2}f_{3,1}(z)\mathcal{K}_{-1/2,-1}(\hat b,z)\,,\nonumber\\
\mathcal{S}_2(z)&=& -\frac{1}{2\hat{b}^4}\Big[\hat{b}^2f_{2,2}(z)+\left(4-3 \hat{b}^2\right) f_{4,2}(z)+6 \hat{b}f_{5,2}(z)+ 2 \hat{b}^2 f_{6,2}(z)-8 \hat{b} f_{7,2}(z)+ 8f_{8,2}(z)\Big]\mathcal{K}_{-3/2,-2}(\hat b,z)\,,\nonumber\\
\mathcal{S}_3(z)&=&-\frac{1}{\hat{b}^6 }\Big[3 \hat{b}^2f_{5,3}(z)-\left(5 \hat{b}^2+4\right)f_{7,3}(z)+10 \hat{b} f_{8,3}(z)+2 \hat{b}^2 f_{9,3}(z)-8 \hat{b}f_{10,3}(z)+8f_{11,3}(z)\Big]\mathcal{K}_{-5/2,-2}(\hat b,z) \,,\nonumber\\
\mathcal{S}_4(z)&=& \frac{1}{8\hat{b}^9}\Big[3 \hat{b}^5f_{4,4}(z)+(24 \hat{b}^3-18 \hat{b}^5)f_{6,4}(z)+36 \hat{b}^4 f_{7,4}(z)+(35 \hat{b}^5-24 \hat{b}^3-16 \hat{b})f_{8,4}(z)+(48 \hat{b}^2-140 \hat{b}^4)f_{9,4}(z)\nonumber\\
&+&(-28 \hat{b}^5+140 \hat{b}^3-64 \hat{b})f_{10,4}(z)+(168 \hat{b}^4+128)f_{11,4}(z)+(8 \hat{b}^5-336 \hat{b}^3)f_{12,4}(z)
+(224 \hat{b}^2-64 \hat{b}^4)f_{13,4}(z)\nonumber\\
&+&192 \hat{b}^3f_{14,4}(z)-256 \hat{b}^2 f_{15,4}(z)+128 \hat{b}f_{16,4}(z)\Big]\mathcal{K}_{-7/2,-3}(\hat b,z)\,,
\eea
\end{widetext}
having denoted
\bea
\mathcal{K}_{m,n}(\hat b,z)&=&\left(1+\frac{2z^3}{\hat{b}(1-z^2)}\right)^m\left(1-\frac{2z}{\hat{b}}\right)^n\nonumber\\
&=& \left(1+\frac{2}{\hat{b}}f_{3,1}(z)\right)^m\left(1-\frac{2}{\hat{b}}f_{1,0}(z)\right)^n\,,
\eea
and
\beq
f_{a,b}(z)=\frac{z^a}{(1-z^2)^b}\,,
\eeq
with the group property 
\beq
f_{a_1,b_1}(z)\cdot f_{a_2,b_2}(z)\cdot\ldots=f_{a_1+a_2+\ldots,b_1+b_2+\ldots}(z)\,.
\eeq
Recalling the binomial relation
\bea
\left(1+\frac{c_1}{\hat{b}}\right)^n&=& \sum_{k=0}^\infty {n \choose k} \frac{c_1^k}{\hat{b}^k}\,,
\eea
we find
\bea
\mathcal{K}_{m,n} &=& \sum_{k,j=0}^\infty {m \choose k}{n \choose j} \left(\frac{2 }{\hat{b}}\right)^{k+j} \frac{z^{3k+j}}{(1-z^2)^k}\nonumber\\
&=& \sum_{k,j=0}^\infty {m \choose k}{n \choose j} \left(\frac{2 }{\hat{b}}\right)^{k+j} f_{3k+j,k}(z)\,,
\eea
so that the functions \eqref{Sn} can be all expressed as (infinite) sums of terms proportional to $f_{3k+j+a,k+b}(z)$.
For example, the term linear in $\hat a$ reads
\beq
\mathcal{S}_1(z)=-\frac{2}{\hat{b}^2}\sum_{k,j=0}^\infty {-\frac12 \choose k}{-1 \choose j} \left(\frac{2 }{\hat{b}}\right)^{k+j}
f_{3k+j+3,k+1}(z)\,.
\eeq
To obtain the radial action one have to multiply the functions $\mathcal{S}_k(z)$ by the prefactor $\sqrt{1-z^2}/z^2=f_{-2,-\frac12}(z)$ (see Eq. \eqref{conprefac}), which again shifts the indices $i,j$ in the coefficients $f_{i,j}(z)$.
Integration over $z$ is then carried out by using the general relation (valid for $a>-1$, $b<1$)
\beq
\int_0^1 dz f_{a,b}(z)  
=\frac12 \frac{ \Gamma\left[\frac{ 1 + a }{2}\right] \Gamma \left[1 - b\right] }{\Gamma \left[\frac{1+a}{2}+(1-b)\right] }\,.
\eeq
Performing the summations, the  radial action finally reads
\bea
I_r&=&ME \sum_{n=0}^\infty I_r^{(n)}{\hat a}^n\nonumber\\
&=&ME\Big[I_r^{(0)}{+}I_r^{(1)}{\hat a} {+}I_r^{(2)}{\hat a}^2 {+} I_r^{(3)}{\hat a}^3{+} I_r^{(4)}{\hat a}^4+O({\hat a}^5)\Big]\,.\nonumber\\
\eea
The various $I_r^{(n)}\, (n=0,\ldots 4)$ as functions of $x=1/\hat b$ are listed below in Tables \ref{tab:1} and \ref{tab:1bis}, once decomposed in a $\pi$-part ($I_r^{\pi (n)}$) and a non-$\pi$ ($I_r^{\slashed{\pi} (n)}$) part (or, equivalently, in $x$-odd and $x$-even parts.)
Notice that having performed all the nontrivial summations the final result is exact in $\hat b$, i.e., PM-exact.

\begin{widetext}
\begin{table*}  
\caption{\label{tab:1}  Contributions to the the radial action corresponding to various powers of $\hat a$. Here (and below) $x=\frac{1}{\hat b}$.}
\begin{ruledtabular}
\begin{tabular}{ll}
$I_r^{\pi (0)}$ & $ -\frac{1}{2x} \pi    \,
   _3F_2\left(-\frac{1}{2},\frac{1}{6},\frac{5}{6
   };\frac{1}{2},1;27x^2\right)$\\
&$\sim -\pi\left(\frac{1}{2x}-\frac{15}{8}x-\frac{1155}{128}x^3 +O(x^5)\right)$\\
$I_r^{\slashed{\pi}(0)}$ & $\frac{32 x^2}{3} \,
   _4F_3\left(1,1,\frac{5}{3},\frac{7}{3};2,\frac
   {5}{2},\frac{5}{2};27x^2\right)
   $\\
&$\sim\frac{32}{3}x^2+\frac{448}{5}x^4+\frac{8192}{7}x^6+O(x^8)$\\
\hline
$I_r^{\pi (1)}$ & $-\frac{5x^2 \pi}{8}  \left(3 \,
   _3F_2\left(\frac{7}{6},\frac{3}{2},\frac{11}{6
   };\frac{5}{2},3;27x^2\right)+\,
   _4F_3\left(\frac{7}{6},\frac{3}{2},\frac{3}{2}
   ,\frac{11}{6};\frac{1}{2},\frac{5}{2},3;27x^2\right)\right)$\\
&$\sim-\pi \left(\frac52 x^2+\frac{693}{16}x^4+\frac{109395}{128}x^6+O(x^8) \right)$\\
$I_r^{\slashed{\pi}(1)}$ & $-\big[\frac{2x}{3 } \,
   _4F_3\left(\frac{2}{3},1,1,\frac{4}{3};\frac{1
   }{2},2,\frac{5}{2};27x^2\right)
   -\frac{1}{3x}  \left(1-\,
   _3F_2\left(-\frac{1}{3},\frac{1}{3},1;-\frac{1
   }{2},\frac{3}{2};27x^2\right)\right)\big]$\\
&$\sim -2x-32x^3-\frac{1792}{3}x^5+O(x^7)$\\
\hline
$I_r^{\pi (2)}$ & $-\big[-\frac{1}{8} x \left(63 x^2 \,
   _3F_2\left(\frac{11}{6},\frac{13}{6},\frac{5}{2};2,\frac{7}{2};27
   x^2\right)+\left(8 x^2-6\right) \,
   _4F_3\left(\frac{7}{6},\frac{3}{2},\frac{3}{2},\frac{11}{6};\frac{1}{2},1
   ,\frac{7}{2};27 x^2\right)+6 \left(x^2 \left(27 \,
   _4F_3\left(\frac{11}{6},\frac{13}{6},\frac{5}{2},\frac{5}{2};\frac{3}{2},
   2,\frac{9}{2};27 x^2\right)\right.\right.\right.$\\ 
&$\left.\left.\left. +5 \,
   _4F_3\left(\frac{11}{6},\frac{13}{6},\frac{5}{2},\frac{7}{2};\frac{1}{2},
   3,\frac{11}{2};27 x^2\right)-55 \,
   _4F_3\left(\frac{13}{6},\frac{5}{2},\frac{17}{6},\frac{7}{2};\frac{3}{2},
   3,\frac{11}{2};27 x^2\right)\right)+\,
   _4F_3\left(\frac{3}{2},\frac{11}{6},\frac{13}{6},\frac{5}{2};\frac{1}{2},
   2,\frac{9}{2};27 x^2\right)\right)\right)\big]$\\
&$\sim \pi\left(\frac{95}{32}x^3+\frac{15939}{128}x^5+\frac{16518645}{4096}x^7+O(x^9)\right)$\\
$I_r^{\slashed{\pi}(2)}$ & $\frac{1}{3072 x^6}\Big[ -192 x^2 \,
   _4F_3\left(-\frac{5}{3},-\frac{4}{3},1,1;-\frac{5}{2},-\frac{1}{2},2;27
   x^2\right)+160 x^2 \,
   _4F_3\left(-\frac{4}{3},-\frac{2}{3},1,1;-\frac{5}{2},-\frac{1}{2},2;27
   x^2\right)$\\
&$-64 x^2 \,
   _4F_3\left(-\frac{2}{3},-\frac{1}{3},1,1;-\frac{3}{2},-\frac{1}{2},2;27
   x^2\right)+\,
   _3F_2\left(-\frac{8}{3},-\frac{7}{3},1;-\frac{7}{2},-\frac{3}{2};27
   x^2\right)-3072 x^8 \,
   _4F_3\left(\frac{2}{3},1,1,\frac{4}{3};\frac{1}{2},\frac{1}{2},3;27
   x^2\right)$\\
&$-512 x^6 \,
   _4F_3\left(-\frac{1}{3},\frac{1}{3},1,1;-\frac{1}{2},-\frac{1}{2},2;27
   x^2\right)-768 x^6 \,
   _4F_3\left(\frac{1}{3},\frac{2}{3},1,1;-\frac{1}{2},\frac{1}{2},2;27
   x^2\right)+2304 x^6 \,
   _4F_3\left(\frac{1}{3},\frac{2}{3},1,1;-\frac{1}{2},\frac{1}{2},3;27
   x^2\right)$\\
&$+2304 x^6 \,
   _4F_3\left(\frac{2}{3},1,1,\frac{4}{3};\frac{1}{2},\frac{1}{2},3;27
   x^2\right)-576 x^4 \,
   _4F_3\left(-\frac{2}{3},-\frac{1}{3},1,1;-\frac{3}{2},-\frac{1}{2},2;27
   x^2\right)+4608 x^4 \,
   _4F_3\left(-\frac{2}{3},-\frac{1}{3},1,1;-\frac{3}{2},\frac{1}{2},3;27
   x^2\right)$\\
&$-3072 x^4 \,
   _4F_3\left(-\frac{1}{3},\frac{1}{3},1,1;-\frac{3}{2},\frac{1}{2},3;27
   x^2\right)+384 x^4 \,
   _4F_3\left(-\frac{1}{3},\frac{1}{3},1,1;-\frac{1}{2},-\frac{1}{2},2;27
   x^2\right)+512 x^4 \,
   _4F_3\left(\frac{1}{3},\frac{2}{3},1,1;-\frac{1}{2},\frac{1}{2},3;27
   x^2\right)$\\
&$+1280 x^6-64 x^4+64 x^2-1\big]$\\
&$\sim x^2+64 x^4+\frac{6784}{3}x^6+O(x^8)$\\
\hline
$I_r^{\pi (3)}$ & $-\frac{1}{8} x^4 \Big(-108 \,
   _4F_3\left(\frac{11}{6},\frac{13}{6},\frac{5}{2},\frac{5}{2};1,\frac{3}{2
   },\frac{9}{2};27 x^2\right)-5 \left(630 x^2 \,
   _4F_3\left(\frac{17}{6},\frac{19}{6},\frac{7}{2},\frac{11}{2};\frac{3}{2}
   ,4,\frac{15}{2};27 x^2\right)+20 \,
   _4F_3\left(\frac{11}{6},\frac{13}{6},\frac{5}{2},\frac{7}{2};\frac{1}{2},
   2,\frac{11}{2};27 x^2\right)\right.$\\
&$\left. -22 \left(4 x^2+5\right) \,
   _4F_3\left(\frac{13}{6},\frac{5}{2},\frac{17}{6},\frac{7}{2};\frac{3}{2},
   2,\frac{11}{2};27 x^2\right)-28 \,
   _4F_3\left(\frac{13}{6},\frac{5}{2},\frac{17}{6},\frac{9}{2};\frac{1}{2},
   3,\frac{13}{2};27 x^2\right)+91 \,
   _4F_3\left(\frac{5}{2},\frac{17}{6},\frac{19}{6},\frac{9}{2};\frac{3}{2},
   3,\frac{13}{2};27 x^2\right)\right)\Big)$\\
&$\sim -\pi \left(\frac{27}{8}x^4+\frac{4455}{16}x^6+\frac{14209195}{1024}x^8+O(x^{10})\right)$\\
$I_r^{\slashed{\pi}(3)}$ & $-\frac{1}{36864 x^7}  \Big[50 x^2 \,
   _3F_2\left(-\frac{8}{3},-\frac{7}{3},1;-\frac{7}{2},-\frac{5}{2};27
   x^2\right)-576 x^2 \,
   _4F_3\left(-\frac{8}{3},-\frac{7}{3},1,1;-\frac{9}{2},-\frac{3}{2},2;27
   x^2\right)$\\
&$+1024 x^2 \,
   _4F_3\left(-\frac{7}{3},-\frac{5}{3},1,1;-\frac{7}{2},-\frac{3}{2},2;27
   x^2\right)-224 x^2 \,
   _4F_3\left(-\frac{5}{3},-\frac{4}{3},1,1;-\frac{7}{2},-\frac{3}{2},2;27
   x^2\right)$\\
&$+3 \,
   _3F_2\left(-\frac{11}{3},-\frac{10}{3},1;-\frac{11}{2},-\frac{5}{2};27
   x^2\right)-6 \,
   _3F_2\left(-\frac{10}{3},-\frac{8}{3},1;-\frac{9}{2},-\frac{5}{2};27
   x^2\right)-12288 x^8 \,
   _4F_3\left(-\frac{1}{3},\frac{1}{3},1,1;-\frac{3}{2},-\frac{1}{2},3;27
   x^2\right)$\\
&$+9216 x^8 \,
   _4F_3\left(\frac{1}{3},\frac{2}{3},1,1;-\frac{1}{2},-\frac{1}{2},3;27
   x^2\right)+1920 x^6 \,
   _4F_3\left(-\frac{4}{3},-\frac{2}{3},1,1;-\frac{5}{2},-\frac{3}{2},2;27
   x^2\right)$\\
&$-6912 x^6 \,
   _4F_3\left(-\frac{2}{3},-\frac{1}{3},1,1;-\frac{3}{2},-\frac{3}{2},2;27
   x^2\right)+46080 x^6 \,
   _4F_3\left(-\frac{2}{3},-\frac{1}{3},1,1;-\frac{3}{2},-\frac{1}{2},3;27
   x^2\right)$\\
&$-15360 x^6 \,
   _4F_3\left(-\frac{1}{3},\frac{1}{3},1,1;-\frac{3}{2},-\frac{1}{2},3;27
   x^2\right)+10752 x^4 \,
   _4F_3\left(-\frac{5}{3},-\frac{4}{3},1,1;-\frac{7}{2},-\frac{1}{2},3;27
   x^2\right)$\\
&$-5760 x^4 \,
   _4F_3\left(-\frac{5}{3},-\frac{4}{3},1,1;-\frac{5}{2},-\frac{3}{2},2;27
   x^2\right)+2400 x^4 \,
   _4F_3\left(-\frac{4}{3},-\frac{2}{3},1,1;-\frac{5}{2},-\frac{3}{2},2;27
   x^2\right)$\\
&$-15360 x^4 \,
   _4F_3\left(-\frac{4}{3},-\frac{2}{3},1,1;-\frac{5}{2},-\frac{1}{2},3;27
   x^2\right)+3840 x^4 \,
   _4F_3\left(-\frac{2}{3},-\frac{1}{3},1,1;-\frac{5}{2},-\frac{1}{2},3;27
   x^2\right)$\\
&$+6144 x^8-8832 x^6+2144 x^4-218 x^2+3\Big]$\\
&$\sim -\frac23 x^3 -\frac{320}{3}x^5-6400 x^7+O(x^9)$\\
 \end{tabular}
\end{ruledtabular}
\end{table*}

\begin{table*}  
\caption{\label{tab:1bis}  Contributions to the the radial action corresponding to   $\hat a^4$}
\begin{ruledtabular}
\begin{tabular}{ll}
$I_r^{\pi (4)}$ & $-\frac{1}{128} x^3 \Big[3744 x^2 \,
   _4F_3\left(\frac{5}{2},\frac{17}{6},\frac{19}{6},\frac{7}{2};1,\frac{3}{2
   },\frac{13}{2};27 x^2\right)+3920 x^2 \,
   _4F_3\left(\frac{5}{2},\frac{17}{6},\frac{19}{6},\frac{9}{2};\frac{1}{2},
   2,\frac{15}{2};27 x^2\right)$\\
&$-14040 x^2 \,
   _4F_3\left(\frac{17}{6},\frac{19}{6},\frac{7}{2},\frac{7}{2};\frac{3}{2},
   2,\frac{13}{2};27 x^2\right)-58800 x^2 \,
   _4F_3\left(\frac{17}{6},\frac{19}{6},\frac{7}{2},\frac{9}{2};\frac{3}{2},
   2,\frac{15}{2};27 x^2\right)$\\
&$-21168 x^2 \,
   _4F_3\left(\frac{17}{6},\frac{19}{6},\frac{7}{2},\frac{11}{2};\frac{1}{2}
   ,3,\frac{17}{2};27 x^2\right)+179928 x^2 \,
   _4F_3\left(\frac{19}{6},\frac{7}{2},\frac{23}{6},\frac{11}{2};\frac{3}{2}
   ,3,\frac{17}{2};27 x^2\right)$\\
&$+22176 x^2 \,
   _4F_3\left(\frac{19}{6},\frac{7}{2},\frac{23}{6},\frac{13}{2};\frac{1}{2}
   ,4,\frac{19}{2};27 x^2\right)-140448 x^2 \,
   _4F_3\left(\frac{7}{2},\frac{23}{6},\frac{25}{6},\frac{13}{2};\frac{3}{2}
   ,4,\frac{19}{2};27 x^2\right)$\\
&$-784 \,
   _4F_3\left(\frac{5}{2},\frac{17}{6},\frac{19}{6},\frac{9}{2};\frac{1}{2},
   2,\frac{15}{2};27 x^2\right)+504 \,
   _4F_3\left(\frac{17}{6},\frac{19}{6},\frac{7}{2},\frac{11}{2};\frac{1}{2}
   ,3,\frac{17}{2};27 x^2\right)+137088 x^6 \,
   _4F_3\left(\frac{19}{6},\frac{7}{2},\frac{23}{6},\frac{11}{2};\frac{3}{2}
   ,3,\frac{17}{2};27 x^2\right)$\\
&$-1792 x^4 \,
   _4F_3\left(\frac{5}{2},\frac{17}{6},\frac{19}{6},\frac{9}{2};\frac{1}{2},
   2,\frac{15}{2};27 x^2\right)+18720 x^4 \,
   _4F_3\left(\frac{17}{6},\frac{19}{6},\frac{7}{2},\frac{7}{2};\frac{3}{2},
   2,\frac{13}{2};27 x^2\right)$\\
&$+20160 x^4 \,
   _4F_3\left(\frac{17}{6},\frac{19}{6},\frac{7}{2},\frac{9}{2};\frac{3}{2},
   2,\frac{15}{2};27 x^2\right)+62985 x^4 \,
   _4F_3\left(\frac{7}{2},\frac{7}{2},\frac{23}{6},\frac{25}{6};\frac{5}{2},
   3,\frac{13}{2};27 x^2\right)$\\
&$+491568 x^4 \,
   _4F_3\left(\frac{7}{2},\frac{23}{6},\frac{25}{6},\frac{13}{2};\frac{3}{2}
   ,4,\frac{19}{2};27 x^2\right)+24024 x^4 \,
   _4F_3\left(\frac{7}{2},\frac{23}{6},\frac{25}{6},\frac{15}{2};\frac{1}{2}
   ,5,\frac{21}{2};27 x^2\right)$\\
&$-1009008 x^4 \,
   _4F_3\left(\frac{23}{6},\frac{25}{6},\frac{9}{2},\frac{15}{2};\frac{3}{2}
   ,5,\frac{21}{2};27 x^2\right)-8 \left(16 x^4+24 x^2-35\right) \,
   _4F_3\left(\frac{13}{6},\frac{5}{2},\frac{17}{6},\frac{7}{2};\frac{1}{2},
   1,\frac{13}{2};27 x^2\right)\Big]$\\
&$\sim \pi \left(\frac{239}{64}x^5+\frac{1094445}{2048}x^7+\frac{159534375}{4096}x^9+O(x^{11})\right)$\\
$I_r^{\slashed{\pi}(4)}$ & $-\big\{\frac{1}{360} \left(36 \,
   _4F_3\left(-\frac{1}{3},\frac{1}{3},1,1;-\frac{3}{2},-\frac{3}{2},4;27
   x^2\right)-36 \,
   _4F_3\left(\frac{1}{3},\frac{2}{3},1,1;-\frac{3}{2},-\frac{1}{2},4;27
   x^2\right)-9 \,
   _4F_3\left(1,1,\frac{4}{3},\frac{5}{3};-\frac{3}{2},\frac{1}{2},4;27
   x^2\right)+20\right) x^2 $\\
&$ +\frac{1}{1920}\Big(-1120 \,
   _4F_3\left(-\frac{4}{3},-\frac{2}{3},1,1;-\frac{7}{2},-\frac{3}{2},4;27
   x^2\right)-150 \,
   _4F_3\left(-\frac{4}{3},-\frac{2}{3},1,1;-\frac{5}{2},-\frac{5}{2},3;27
   x^2\right)$\\
&$+720 \,
   _4F_3\left(-\frac{2}{3},-\frac{1}{3},1,1;-\frac{5}{2},-\frac{3}{2},3;27
   x^2\right)-80 \,
   _4F_3\left(-\frac{2}{3},-\frac{1}{3},1,1;-\frac{5}{2},-\frac{3}{2},4;27
   x^2\right)+288 \,
   _4F_3\left(-\frac{1}{3},\frac{1}{3},1,1;-\frac{3}{2},-\frac{3}{2},4;27
   x^2\right)$\\
&$+30 \,
   _4F_3\left(\frac{1}{3},\frac{2}{3},1,1;-\frac{5}{2},-\frac{1}{2},3;27
   x^2\right)-144 \,
   _4F_3\left(\frac{1}{3},\frac{2}{3},1,1;-\frac{3}{2},-\frac{3}{2},4;27
   x^2\right)+144 \,
   _4F_3\left(\frac{1}{3},\frac{2}{3},1,1;-\frac{3}{2},-\frac{1}{2},4;27
   x^2\right)-109\Big)$\\
&$+\frac{1}{2304 x^2}\Big(288 \,
   _4F_3\left(-\frac{7}{3},-\frac{5}{3},1,1;-\frac{9}{2},-\frac{5}{2},3;27x^2\right)+7 \,
   _4F_3\left(-\frac{7}{3},-\frac{5}{3},1,1;-\frac{7}{2},-\frac{7}{2},2;27 x^2\right)$\\
&$-63 \,
   _4F_3\left(-\frac{5}{3},-\frac{4}{3},1,1;-\frac{7}{2},-\frac{5}{2},2;27x^2\right)+154 \,
   _4F_3\left(-\frac{5}{3},-\frac{4}{3},1,1;-\frac{7}{2},-\frac{5}{2},3;27x^2\right)-270 \,
   _4F_3\left(-\frac{4}{3},-\frac{2}{3},1,1;-\frac{5}{2},-\frac{5}{2},3;27x^2\right)$\\
&$+270 \,
   _4F_3\left(-\frac{2}{3},-\frac{1}{3},1,1;-\frac{5}{2},-\frac{5}{2},3;27x^2\right)-648 \,
   _4F_3\left(-\frac{2}{3},-\frac{1}{3},1,1;-\frac{5}{2},-\frac{3}{2},3;27x^2\right)+840 \,
   _4F_3\left(-\frac{2}{3},-\frac{1}{3},1,1;-\frac{5}{2},-\frac{3}{2},4;27x^2\right)$\\
&$-504 \,
   _4F_3\left(-\frac{1}{3},\frac{1}{3},1,1;-\frac{3}{2},-\frac{3}{2},4;27x^2\right)+486\Big)$\\
&$+\frac{1}{184320 x^4}\Big(-462 \,
   _4F_3\left(-\frac{10}{3},-\frac{8}{3},1,1;-\frac{11}{2},-\frac{7}{2},2;27x^2\right)-525 \,
   _4F_3\left(-\frac{8}{3},-\frac{7}{3},1,1;-\frac{9}{2},-\frac{7}{2},2;27x^2\right)$\\
&$+840 \,
   _4F_3\left(-\frac{7}{3},-\frac{5}{3},1,1;-\frac{7}{2},-\frac{7}{2},2;27x^2\right)-94080 \,
   _4F_3\left(-\frac{5}{3},-\frac{4}{3},1,1;-\frac{9}{2},-\frac{3}{2},4;27x^2\right)$\\
&$+3780 \,
   _4F_3\left(-\frac{5}{3},-\frac{4}{3},1,1;-\frac{7}{2},-\frac{5}{2},2;27x^2\right)-49000 \,
   _4F_3\left(-\frac{5}{3},-\frac{4}{3},1,1;-\frac{7}{2},-\frac{5}{2},3;27x^2\right)$\\
&$+282240 \,
   _4F_3\left(-\frac{5}{3},-\frac{4}{3},1,1;-\frac{7}{2},-\frac{3}{2},4;27x^2\right)-141120 \,
   _4F_3\left(-\frac{4}{3},-\frac{2}{3},1,1;-\frac{7}{2},-\frac{3}{2},4;27x^2\right)$\\
&$+31500 \,
   _4F_3\left(-\frac{4}{3},-\frac{2}{3},1,1;-\frac{5}{2},-\frac{5}{2},3;27x^2\right)+26880 \,
   _4F_3\left(-\frac{2}{3},-\frac{1}{3},1,1;-\frac{5}{2},-\frac{3}{2},4;27x^2\right)+10117\Big)$\\
&$+\frac{1}{73728 x^6}\Big(4620 \,
   _4F_3\left(-\frac{8}{3},-\frac{7}{3},1,1;-\frac{11}{2},-\frac{5}{2},3;27x^2\right)+17952 \,
   _4F_3\left(-\frac{8}{3},-\frac{7}{3},1,1;-\frac{11}{2},-\frac{3}{2},4;27x^2\right)$\\
&$+735 \,
   _4F_3\left(-\frac{8}{3},-\frac{7}{3},1,1;-\frac{9}{2},-\frac{7}{2},2;27x^2\right)-22680 \,
   _4F_3\left(-\frac{8}{3},-\frac{7}{3},1,1;-\frac{9}{2},-\frac{5}{2},3;27x^2\right)$\\
&$+12096 \,
   _4F_3\left(-\frac{7}{3},-\frac{5}{3},1,1;-\frac{9}{2},-\frac{5}{2},3;27x^2\right)-27648 \,
   _4F_3\left(-\frac{7}{3},-\frac{5}{3},1,1;-\frac{9}{2},-\frac{3}{2},4;27x^2\right)$\\
&$-490 \,
   _4F_3\left(-\frac{7}{3},-\frac{5}{3},1,1;-\frac{7}{2},-\frac{7}{2},2;27x^2\right)+10752 \,
   _4F_3\left(-\frac{5}{3},-\frac{4}{3},1,1;-\frac{9}{2},-\frac{3}{2},4;27x^2\right)$\\
&$-3136 \,
   _4F_3\left(-\frac{5}{3},-\frac{4}{3},1,1;-\frac{7}{2},-\frac{5}{2},3;27x^2\right)-2688 \,
   _4F_3\left(-\frac{5}{3},-\frac{4}{3},1,1;-\frac{7}{2},-\frac{3}{2},4;27x^2\right)+5523\Big)$\\
&$+\frac{1}{1474560 x^8}\Big(-1274 \,_4F_3\left(-\frac{11}{3},-\frac{10}{3},1,1;-\frac{13}{2},-\frac{7}{2},2;27 x^2\right)-43056 \,
   _4F_3\left(-\frac{11}{3},-\frac{10}{3},1,1;-\frac{13}{2},-\frac{5}{2},3;27 x^2\right)$\\
&$+13230 \,
   _4F_3\left(-\frac{11}{3},-\frac{10}{3},1,1;-\frac{11}{2},-\frac{7}{2},2;27 x^2\right)-4851 \,
   _4F_3\left(-\frac{10}{3},-\frac{8}{3},1,1;-\frac{11}{2},-\frac{7}{2},2;27x^2\right)$\\
&$+71280 \,
   _4F_3\left(-\frac{10}{3},-\frac{8}{3},1,1;-\frac{11}{2},-\frac{5}{2},3;27x^2\right)-26400 \,
   _4F_3\left(-\frac{8}{3},-\frac{7}{3},1,1;-\frac{11}{2},-\frac{5}{2},3;27x^2\right)$\\
&$+1470 \,
   _4F_3\left(-\frac{8}{3},-\frac{7}{3},1,1;-\frac{9}{2},-\frac{7}{2},2;27x^2\right)+10800 \,
   _4F_3\left(-\frac{8}{3},-\frac{7}{3},1,1;-\frac{9}{2},-\frac{5}{2},3;27x^2\right)-15953\Big)$\\
&$+\frac{1}{5898240 x^{10}}\Big(-189 \,
   _3F_2\left(-\frac{14}{3},-\frac{13}{3},1;-\frac{13}{2},-\frac{9}{2};27x^2\right)+3480 \,
   _4F_3\left(-\frac{14}{3},-\frac{13}{3},1,1;-\frac{15}{2},-\frac{7}{2},2;27 x^2\right)$\\
&$-6048 \,
   _4F_3\left(-\frac{13}{3},-\frac{11}{3},1,1;-\frac{13}{2},-\frac{7}{2},2;27 x^2\right)+1456 \,
   _4F_3\left(-\frac{11}{3},-\frac{10}{3},1,1;-\frac{13}{2},-\frac{7}{2},2;27 x^2\right)$\\
&$-1260 \,
   _4F_3\left(-\frac{11}{3},-\frac{10}{3},1,1;-\frac{11}{2},-\frac{7}{2},2;27 x^2\right)+2341\Big)$\\
&$+\frac{1}{15728640 x^{12}}\Big(-25 \,
   _3F_2\left(-\frac{17}{3},-\frac{16}{3},1;-\frac{17}{2},-\frac{9}{2};27x^2\right)+45 \,
   _3F_2\left(-\frac{16}{3},-\frac{14}{3},1;-\frac{15}{2},-\frac{9}{2};27x^2\right)$\\
&$+12 \,
   _3F_2\left(-\frac{14}{3},-\frac{13}{3},1;-\frac{13}{2},-\frac{9}{2};27x^2\right)-32\Big)\big\}$\\
&$\sim \frac12 x^4 +160 x^6 +\frac{45248}{3}x^8+O(x^{10})$\\
 \end{tabular}
\end{ruledtabular}
\end{table*}

\end{widetext}

From the radial action the scattering angle is simply obtained via the relation
\beq
\frac{\pi+\chi}{2}=-\frac{\partial I_r}{\partial J}\,.
\eeq

The expanded form of the scattering angle for large values of the impact parameter is then
\begin{widetext}
\bea
\chi&=&\frac{\frac{15 \pi }{4}-4 {\hat a}}{\hat{b}^2}+\frac{4 {\hat a} ^2-10 \pi  {\hat a} +\frac{128}{3}}{\hat{b}^3}+\frac{-4 {\hat a} ^3+\frac{285 \pi  {\hat a} ^2}{16}-192
   {\hat a} +\frac{3465 \pi }{64}}{\hat{b}^4}\nonumber\\
   &+&\frac{4 {\hat a} ^4-27 \pi  {\hat a} ^3+512 {\hat a}
   ^2-\frac{693 \pi  {\hat a} }{2}+\frac{3584}{5}}{\hat{b}^5}+\frac{\frac{1195 \pi  {\hat a} ^4}{32}-\frac{3200 {\hat a}
   ^3}{3}+\frac{79695 \pi  {\hat a}
   ^2}{64}-\frac{17920 {\hat a} }{3}+\frac{255255
   \pi }{256}}{\hat{b}^6}\nonumber\\
   &+&\frac{1920 {\hat a} ^4-\frac{13365 \pi  {\hat a}
   ^3}{4}+27136 {\hat a} ^2-\frac{328185 \pi 
   {\hat a} }{32}+\frac{98304}{7}}{\hat{b}^7}+O\left({\hat a}^5,\frac{1}{\hat{b}^{8}}\right)\,\nonumber\\
&=& \sum_{k=0}^4\chi^{(k)}+O\left({\hat a}^5,\frac{1}{\hat{b}^{8}}\right)\,.
\eea
For example, using the notation $x=\frac{1}{\hat b}$ and distinguishing among $\pi$ and $\slashed{\pi}$ contributions, the above relation implies
\bea
\chi^{(0)}&=& \pi \left(\frac{15 }{4}x^2+\frac{3465 }{64 }x^4+\frac{255255 }{256}x^6\right) +\frac{128}{3}x^3+\frac{3584}{5}x^5
   +\frac{98304}{7}x^7+O(x^8)\,,\nonumber\\
\chi^{(1)}&=& \pi\left(- 10  x^3 -\frac{693 }{2}x^5-\frac{328185 }{32}x^7\right)- 4x^2- 192x^4 -\frac{17920 }{3}x^6+O(x^8)\,,\nonumber\\
\chi^{(2)}&=& \pi\left( \frac{285}{16}x^4 +\frac{79695}{64}\right)+ 4x^3+512 x^5+27136 x^7+ O(x^8)\,,\nonumber\\
\chi^{(3)}&=& \pi\left( -27 x^5 -\frac{13365}{4}x^7\right)-4x^4-\frac{3200}{3}x^6 + O(x^8)\,,\nonumber\\
\chi^{(4)}&=& \pi \frac{1195}{32}x^6  +4x^5 +1920 x^7+ O(x^8)\,, 
\eea
etc. 
\end{widetext}

\section{Scalar wave equation}

Let us consider the source-free massless scalar wave equation in the Kerr spacetime
\beq
\label{box_psi}
\Box \Psi=0\,,
\eeq
which is separable in the frequency domain as follows
\beq
\Psi(t,r,\theta,\phi)=\sum_{lm}\int \frac{d\omega}{2\pi}e^{-i\omega t+i m\phi}R_{lm\omega}(r)S_{lm\omega}(\theta)\,,
\eeq
in terms of spheroidal harmonics $S_{lm\omega}(\theta)$.
The latter satisfy the equation
\bea
0&=&\left\{\frac{1}{\sin\theta}\frac{d}{d\theta} \left( \sin\theta \frac{d}{d\theta} \right)\right.\nonumber\\
&+&\left.
\left[\xi^2 \cos^2\theta -\frac{m^2}{\sin^2\theta}+E_{(l,m;\xi)}\right]\right\}S_{lm\omega}(\theta)
\,,
\eea
with eigenvalues $E_{(l,m;\xi)}$ and $\xi=a\omega$.
The radial equation coincides with the spin-$0$ homogeneous Teukolsky equation~\cite{Press:1973zz}
\bea
\label{eqrad}
\left\{\frac{d}{dr} \left(\Delta\frac{d}{dr} \right) +\left[\frac{K^2}{\Delta} -\lambda_{(l,m;\xi)}\right]\right\}R_{lm\omega }(r)&=&0\,,
\eea
where $K=(r^2+a^2)\omega -m a$ and 
\beq
\label{lambda_def}
\lambda_{(l,m;\xi)}=E_{(l,m;\xi)}-2m\xi +\xi^2\,,
\eeq
with
\bea
E_{(l,m;\xi)}&=& l(l + 1) 
-\frac{2l^2+2l-2m^2-1}{(2l-1)(2l+3)}\xi^2 \nonumber\\ 
&-&\left[\frac{(4m^2-9)(1-4m^2)}{128(2l+5)(2l-3)} +\frac{16m^4-40m^2-7}{128(2l-1)(2l+3)}\right.\nonumber\\ 
&+&\left.
\frac{(1-4m^2)^2(12l^2+12l+7)}{32(2l+3)^3(2l-1)^3} \right] \xi^4 \nonumber\\
&+&O(\xi^6)\,.
\eea

Let us consider an incoming monocromatic plane wave (PW) with given frequency $\omega$ scattered off the black hole.
According to the standard partial wave expansion (see, e.g., Refs. \cite{Futterman:1988ni,Andersson:1995vi,Dolan:2008kf,Glampedakis:2001cx,Bautista:2021wfy}) the total field at spatial infinity ($r\to\infty$) will be the sum of a plane wave plus an outgoing scattered wave, i.e., 
\beq
\Psi=\Psi_{\rm PW}+\frac{f(\theta,\phi)}{r}e^{i\omega r_*}\,, 
\eeq
where \cite{Bautista:2021wfy}
\beq
f(\theta,\phi)=\frac{2\pi}{i\omega}\sum_{lm}S_{lm\omega}\left(\frac{\pi}{2}\right)S_{lm\omega}(\theta)e^{i m\phi}\left(e^{2i\delta_{lm\omega}}-1\right)\,,
\eeq
denotes the scattering amplitude (for incidence orthogonal to the symmetry axis).

The solution to Eq. \eqref{eqrad} which describes waves that are purely ingoing at the horizon is given by (see, e.g., Sec. 2.1, Eqs. (19) of Ref. \cite{Sasaki:2003xr})
\bea
R_{lm\omega}^{\rm in}&\sim& e^{-ikr_*}\,, \quad r\to r_+\,,\nonumber\\
R_{lm\omega}^{\rm in}&\sim& B_{lm\omega}^{\rm inc}\frac{e^{-i\omega r_*}}{r}+B_{lm\omega}^{\rm ref}\frac{e^{i\omega r_*}}{r}\,, \quad r\to r_\infty\,,\qquad
\eea 
with $k=\omega -ma/2Mr_+$, $r_*=\int \frac{r^2+a^2}{\Delta}dr$ is the tortoise coordinate, and $B_{lm\omega}^{\rm inc}$, $B_{lm\omega}^{\rm ref}$ are the incidence and reflection coefficients, respectively.
The latter are connected to the scattering phase shift $\delta_{lm\omega}$ by the following relation
\bea
\label{deltadef}
e^{2i\delta_{lm\omega}}=(-1)^{l+1}\frac{B_{lm\omega}^{\rm ref}}{B_{lm\omega}^{\rm inc}}\,.
\eea
We will compute below the phase shift in the eikonal limit (i.e., for high frequencies) by using the analytic solutions to the radial homogeneous Teukolsky equation provided by the MST method.

Let us recall that such a regime corresponds to a particle-like behavior of the scattered waves, which propagate along null geodesics.
In a semiclassical approach to scattering \cite{Ford:2000uye} the phase shift is indeed approximated by a one-turning point WKB formula, and is determined by the radial action only in the large-$l$ limit.

\subsection{WKB solutions}

WKB-type solutions \cite{Iyer:1986np,Bini:2015mza} to Eq. \eqref{box_psi} are obtained 1) replacing
\beq
\label{l_omega_defs}
l=\frac{J}{\hbar}-\frac12\,,\qquad \omega=\frac{E}{\hbar}\,,
\eeq
where $\hbar$ is just a small parameter (and has nothing to do with quantum corrections) so  that $\hbar \to 0 \sim l\to \infty$ together with $\omega \to \infty$ but with $\omega/l$ fixed, corresponds to the eikonal limit
and 2) following the ansatz
\beq
\label{S_di_r_gen}
R_{lm\omega}(r)= \frac{e^{ i\sum_{k=0}^\infty \hbar^{k-1}S_k(r)} }{\sqrt{\Delta}}\,   \,,
\eeq
that is, limiting at the LO and NLO levels (i.e., keeping only the terms $O(\frac{1}{\hbar})$ and $O(1)$), 
\beq
\label{S_di_r}
R_{lm\omega}(r)= \frac{e^{\frac{i}{\hbar}S_0(r)+iS_1(r)}}{\sqrt{\Delta}} +O\left(\hbar\right)\,,
\eeq
with $S_0(r)$ simply related to the null geodesics radial momentum $p_r$
\beq
\label{eq_S_0}
\frac1{\hbar^2}\left(\frac{d S_0(r)}{dr} \right)^2 = p_r^2\,,
\eeq
where  $p_r$ (as a function of $r$) was given in Eq. \eqref{separation} above.
Note, in passing, that from Eq. \eqref{l_omega_defs} in the eikonal limit 
\beq
\frac{M\omega}{l}=\frac{M E}{J}=\frac{1}{\hat b}\equiv x\,,
\eeq
and we will work in units of $M$, i.e., assume $M=1$ without any loss of generality.

The WKB function $S_0(r)$ (solution to Eq. \eqref{eq_S_0}), depending explicitly on $r$ and $b=J/E$, is an approximate WKB solution of Eq. \eqref{box_psi}.
When considering the limit $r\to \infty$ of $S_0(r)$, Eq. \eqref{S_di_r}, one is looking at the behavior of the solutions $R_{lm\omega} (r)$ around the (only) irregular singular point of Eq. \eqref{box_psi}.  At the same time,  the limit $\lim_{r\to \infty}S(r)$ gives a function of $b$ only, (i.e., not depending on $r$ anymore). It coincides with the geodesic radial action (here null geodesic radial action),   defined as
\beq
I_r(b)=\int_{r_{\rm min}}^\infty p_r dr\,,
\eeq
with $r_{\rm min}$ denoting the minimum approach distance for null geodesics.

\subsection{MST solutions in the eikonal limit}

Let us turn to the homogeneous radial Teukolsky equation \eqref{eqrad}.
One can apply the MST formalism to find the ingoing and upgoing solutions satisfying the regularity conditions at the outer horizon and at spatial infinity, respectively. 

The incidence and reflection coefficients $B_{lm\omega}^{\rm inc}$ and $B_{lm\omega}^{\rm ref}$ defining the scattering phase shift \eqref{deltadef} can be calculated by using the MST technique (see sec. 4.4, Eqs. (168) and (169) of Ref. \cite{Sasaki:2003xr})  
\begin{widetext}
\bea
B_{lm\omega}^{\rm inc}&=&
\omega^{-1}\left[{K}_{\nu}-
ie^{-i\pi\nu} \frac{\sin \pi(\nu+i\epsilon)}
{\sin \pi(\nu-i\epsilon)}
{K}_{-\nu-1}\right]A_{+}^{\nu} e^{-i(\epsilon\ln\epsilon -\frac{1-\kappa}{2}\epsilon)}\,,\nonumber\\
B_{lm\omega}^{\rm ref}&=&
\omega^{-1}\left[{K}_{\nu}
+ie^{i\pi\nu} {K}_{-\nu-1}\right]A_{-}^{\nu}
e^{i(\epsilon\ln\epsilon -\frac{1-\kappa}{2}\epsilon)}\,,
\eea
with
\bea
A_{+}^\nu&=&e^{-{\pi\over 2}\epsilon}e^{{\pi\over 2}i(\nu+1)}
2^{-1-i\epsilon}{\Gamma(\nu+1+i\epsilon)\over 
\Gamma(\nu+1-i\epsilon)}\sum_{n=-\infty}^{+\infty}a_n^\nu\,,\nonumber\\
A_{-}^\nu&=&e^{-{\pi\over 2}\epsilon}e^{-{\pi\over 2}i(\nu+1)}2^{-1+i\epsilon}
\sum_{n=-\infty}^{+\infty}(-1)^n{(\nu+1-i\epsilon)_n\over 
(\nu+1+i\epsilon)_n}a_n^\nu\,, 
\eea
having denoted the Pochhammer symbols as usual
\beq
(x)_n=\frac{\Gamma(x+n)}{\Gamma(x)}\,,
\eeq
and 
\bea
\label{K_nu_eq}
K_{\nu}&=&	\frac{e^{i\epsilon\kappa}(2\epsilon \kappa )^{-\nu-r}i^{r}
	\Gamma(1-2i\epsilon_+)\Gamma(r+2\nu+2)}
	{\Gamma(r+\nu+1+i\epsilon)
	\Gamma(r+\nu+1+i\tau)\Gamma(r+\nu+1+i\epsilon)}
	\nonumber\\
	&\times& \left ( \sum_{n=r}^{\infty}
	(-1)^n\, \frac{\Gamma(n+r+2\nu+1)}{(n-r)!}
	\frac{\Gamma(n+\nu+1+i\epsilon)}{\Gamma(n+\nu+1-i\epsilon)}
	\frac{\Gamma(n+\nu+1+i\tau)}{\Gamma(n+\nu+1-i\tau)}
	\,a_n^{\nu}\right)
	\nonumber\\
	&\times& \left(\sum_{n=-\infty}^{r}
	\frac{(-1)^n}{(r-n)!
	(r+2\nu+2)_n}\frac{(\nu+1-i\epsilon)_n}{(\nu+1+i\epsilon)_n}
	a_n^{\nu}\right)^{-1}\,.
\eea
\end{widetext}
Here $\epsilon=2 G M \omega$, $\kappa=\sqrt{1-q^2}$, $q=\frac{a}{M}\equiv \hat a$ (preferred standard MST notation in this section to avoid confusion with the recursion coefficients $a_n$), $\tau=\frac{\epsilon-m q}{\kappa}$, $ \epsilon_{\pm}=\frac{\epsilon\pm\tau}{2}$, $r$ is an arbitrary integer number which can be chosen as $r=0$, and $\nu$ is the renormalized angular momentum. The series coefficients $a_n^{\nu}$ satisfy the following three term recurrence relation \cite{Sasaki:2003xr}
\beq
\label{3termrec}
\alpha^\nu_na^\nu_{n+1}+\beta^\nu_na^\nu_n+\gamma^\nu_na^\nu_{n-1}=0\,,
\eeq
where
\bea
\alpha_n^\nu&{=}&\frac{i\epsilon \kappa(n{+}\nu{+}1{+}i\epsilon)(n{+}\nu{+}1{-}i\epsilon)(n{+}\nu{+}1{+}i\tau)}{(n{+}\nu{+}1)(2n{+}2\nu{+}3)}\,,\nonumber\\
\beta_n^\nu&{=}&-\lambda_{(l,m;\xi)}+(n+\nu)(n+\nu+1)+\epsilon(\epsilon- m q)\nonumber\\
&+&\frac{\epsilon^3(\epsilon-m q)}{(n+\nu)(n+\nu+1)}\,,\nonumber\\
\gamma_n^\nu&{=}&{-}\frac{i\epsilon \kappa(n{+}\nu{+}i\epsilon)(n{+}\nu{-}i\epsilon)(n{+}\nu{-}i\tau)}{(n{+}\nu)(2n{+}2\nu{-}1)}\,,
\eea
with the condition $a^\nu_N=0=a^\nu_{-N}$ for a given $n=N$ and $a_0=1$.

It has been suggested in Ref.~\cite{Ivanov:2025ozg} that in the eikonal limit there exists a simple relation between the phase shift $\delta_{lm\omega}$ and the renormalized angular momentum $\nu$ (see Eq. (3) there) 
\beq
\label{fin_res}
 \delta_{lm\omega}+\delta_{lm-\omega} = \pi(\nu-l)\,.
\eeq
In the following we will prove this conjecture in the simplest case of the Schwarzschild spacetime (the Kerr case can be treated in a similar way).

First of all let us note that direct evaluation of the functions $K_\nu$ and $K_{-\nu-1}$ through Eq. \eqref{K_nu_eq} gives the large-$l$ behaviors $K_\nu\sim e^{-l}$ and $K_{-\nu-1}\sim l^le^{l}$, implying that 
\beq
\frac{K_\nu}{K_{-\nu-1}}\sim l^{-l}e^{-2l}\ll 1\,,
\eeq
upon using the solutions for the series coefficients $a_n^{\nu}$ given in Appendix \ref{rec_rel} as functions of $x=\omega/l$ (finite) in the large-$l$ limit.
Therefore, one can neglect the term proportional to $K_\nu$ in Eq. \eqref{deltadef} for the phase shift, which behaves as
\bea
e^{2i\delta_{lm\omega}}&\sim&(-1)^{l+1}S_\nu\frac{A_{-}^{\nu}
}{A_{+}^{\nu}}e^{2i\epsilon\ln\epsilon}\nonumber\\
&=&e^{-i\pi(\nu-l)}S_\nu e^{2i\epsilon\ln(2\epsilon)}{\Gamma(\nu+1-i\epsilon)\over 
\Gamma(\nu+1+i\epsilon)}\nonumber\\
&\times&
\frac{\sum_{n=-\infty}^{+\infty}(-1)^n{(\nu+1-i\epsilon)_n\over 
(\nu+1+i\epsilon)_n}a_n^\nu}{\sum_{n=-\infty}^{+\infty}a_n^\nu}\,,
\eea
where 
\beq
S_\nu(\omega)= - e^{2i\pi \nu}\frac{\sin(\pi (\nu-i\epsilon))}{\sin(\pi (\nu+i\epsilon))}\,.
\eeq
A direct evaluation also implies that
\beq
e^{2i\epsilon\ln(2\epsilon)}{\Gamma(\nu+1-i\epsilon)\over 
\Gamma(\nu+1+i\epsilon)}
\frac{\sum_{n=-\infty}^{+\infty}(-1)^n{(\nu+1-i\epsilon)_n\over 
(\nu+1+i\epsilon)_n}a_n^\nu}{\sum_{n=-\infty}^{+\infty}a_n^\nu}\sim 1\,,
\eeq
[Actually one finds $1+O(x^{N+1})$ if the solution for the recursion coefficients is truncated for $n=N$.]
whence 
\beq
e^{2i\delta_{lm\omega}} 
\sim  e^{-i\pi(\nu-l)}S_\nu \,.
\eeq
Recalling that $\nu(\omega)=\nu(-\omega)$ and $S_\nu(\omega)S_\nu(-\omega)=e^{4i\pi \nu}$ finally leads to
\bea
e^{2i\delta_{lm\omega}} e^{2i\delta_{lm-\omega}} 
&\sim &  e^{-i\pi(\nu-l)}S_\nu(\omega) e^{-i\pi(\nu-l)} S_\nu(-\omega)\nonumber\\
&= &  e^{2i\pi(\nu-l)}\,, 
\eea
which completes the proof of Eq. \eqref{fin_res} in the Schwarzschild case. 

In the Kerr case one proceeds exactly in the same way.
The solution for $\nu$ \footnote{We omit here correction terms of the type $c_2 (\delta_2^l-\delta_2^{-l-1})$ to $\nu_2$, $c_3 (\delta_3^l-\delta_3^{-l-1})$ to $\nu_3$, etc.,  arising when working with fixed values of $l=2,3,\ldots$. See, e.g., Ref. \cite{Bini:2013rfa,Bini:2025ltr}.} in the eikonal limit (with $m=l$ for equatorial scattering) 
  writes 
\beq
\nu=l+ \sum_{k=2}^\infty \nu_k x^k\,,
\eeq
where the various recursion coefficients $a_k$ as well as the renormalized angular momentum $\nu_k$ are listed in Appendix \ref{rec_rel} up to order $O(x^4,q^2)$. The Supplemental Material file associated with this paper contains higher-order expansions.
 
At the leading-order in a large $l$ expansion we have the following relation
\beq
\frac{\Delta \nu}{\nu}\sim \frac{(\nu-l)}{l}\underset{l\to\infty}{\sim}\sum_i G_i \hat a^i\,,
\eeq
where the first four $G_i=G_i(x)$ are listed below in Eq. \eqref{variousG_i} together with their corresponding resummed expressions,
\begin{widetext}
\bea
\label{variousG_i}
G_0(x)&=&-\frac{15}{4}x^2-\frac{1155}{64} x^4-\frac{51051}{256} x^6-\frac{47805615}{16384} x^8-\frac{3234846615}{65536} x^{10}-\frac{957220521075}{1048576} x^{12}+\mathcal{O}(x^{14})\nonumber\\
&=&{}_3F_2(-\frac12, \frac16, \frac56; \frac12, 1; 27 x^2)-1  \,,\nonumber\\
G_1(x)&=&5x^3 + \frac{693}{8}x^5+\frac{109395}{64}x^7+\frac{37182145}{1024}x^9+\frac{13233463425}{16384}x^{11}+O(x^{13})\nonumber\\
&=& 5x^3 {}_3F_2(\frac76,\frac32, \frac{11}{6}; 2, \frac52; 27x^2)\,,\nonumber\\
G_2(x)&=&-\frac{95}{16}x^4 -\frac{15939}{64} x^6-\frac{16518645}{2048} x^8-\frac{1970653685}{8192}x^{10}-\frac{1812984489225}{262144}x^{12}+\mathcal{O}(x^{14})\nonumber\\
&=& \frac{1}{16} x^4 \Big(-11088x^2 \,
   _3F_2\left(\frac{13}{6},\frac{5}{2},\frac{17}{6};2,\frac{7}{2};27 x^2\right)-360 \,
   _3F_2\left(\frac{7}{6},\frac{3}{2},\frac{11}{6};1,\frac{5}{2};27 x^2\right)-5 \,
   _3F_2\left(\frac{7}{6},\frac{3}{2},\frac{11}{6};\frac{5}{2},3;27 x^2\right)\nonumber\\
&&+\frac{5\left(\left(9 x^2-47\right)
   \,
   _2F_1\left(\frac{7}{6},\frac{11}{6};3;27 x^2\right)-7
   \left(189 x^2+1\right) \,
   _2F_1\left(\frac{11}{6},\frac{13}{6};3;27 x^2\right)\right)}{27 x^2-1}\Big)\,,\nonumber\\
G_3(x)&=& \frac{27}{4}x^5 +\frac{4455}{8}x^7+\frac{14209195}{512}x^9+\frac{578013345}{512}x^{11}+O(x^{13})\nonumber\\
&=& \frac{1}{128} x^5 \Big(-31680x^2 \, _3F_2\left(\frac{13}{6},\frac{17}{6},\frac{7}{2};2,\frac{9}{2};27 x^2\right)-1408 \,
   _3F_2\left(\frac{7}{6},\frac{11}{6},\frac{5}{2};1,\frac{7}{2};27 x^2\right)\nonumber\\
&&+2560 \,
   _5F_4\left(\frac{7}{6},\frac{11}{6},2,2,2;1,1,1,3;27x^2\right)-2048 \,
   _5F_4\left(\frac{7}{6},\frac{11}{6},2,2,\frac{5}{2};1,1,1,\frac{7}{2};27x^2\right)\nonumber\\
&&+70840 x^2 \,
   _2F_1\left(\frac{13}{6},\frac{17}{6};4;27x^2\right)+1760 \,
   _2F_1\left(\frac{7}{6},\frac{11}{6};3;27 x^2\right)\nonumber\\
&&+765765 x^4 \,
   _2F_1\left(\frac{19}{6},\frac{23}{6};5;27x^2\right)\Big)\,,
\eea
\bea
G_4(x)&=&-\frac{239}{32} x^6-\frac{1094445}{1024} x^8-\frac{159534375}{2048} x^{10}-\frac{276178832475}{65536} x^{12}+\mathcal{O}(x^{14})\nonumber\\
&=& \frac{x^6 }{12288}
   \Big(3606240 x^2 \,
   _3F_2\left(\frac{13}{6},\frac{17}{6},\frac{7}{2};\frac{9}{2},5;27 x^2\right)+74432\,
   _3F_2\left(\frac{7}{6},\frac{11}{6},\frac{5}{2};\frac{7}{2},4;27 x^2\right)\nonumber\\
&&-136960\,
   _5F_4\left(\frac{7}{6},\frac{11}{6},2,2,2;1,1,1,4;27x^2\right)+657408 \,
   _5F_4\left(\frac{7}{6},\frac{11}{6},2,2,\frac{5}{2};1,1,1,\frac{7}{2};27 x^2\right)\nonumber\\
&&-547840 \,
   _5F_4\left(\frac{7}{6},\frac{11}{6},2,2,3;1,1,1,4;27x^2\right)-143360 \,
   _6F_5\left(\frac{7}{6},\frac{11}{6},2,2,2,2;1,1,1,1,4;27x^2\right)\nonumber\\
&&+688128 \,
   _6F_5\left(\frac{7}{6},\frac{11}{6},2,2,2,\frac{5}{2};1,1,1,1,\frac{7}{2};27x^2\right)-573440 \,
   _6F_5\left(\frac{7}{6},\frac{11}{6},2,2,2,3;1,1,1,1,4;27x^2\right)\nonumber\\
&&-81920 \,
   _7F_6\left(\frac{7}{6},\frac{11}{6},2,2,2,2,2;1,1,1,1,1,4;27 x^2\right)+393216 \,
   _7F_6\left(\frac{7}{6},\frac{11}{6},2,2,2,2,\frac{5}{2}; 1,1,1,1,1,\frac{7}{2};27x^2\right)\nonumber\\
&&-327680 \,
   _7F_6\left(\frac{7}{6},\frac{11}{6},2,2,2,2,3;1,1,1,1,1,4;27 x^2\right)+395482815x^6 \,
   _3F_2\left(\frac{25}{6},\frac{29}{6},\frac{11}{2};\frac{13}{2},7;27x^2\right)\nonumber\\
&&+63065002 x^4 \,
   _3F_2\left(\frac{19}{6},\frac{23}{6},\frac{9}{2};\frac{11}{2},6;27x^2\right)-3756060 x^2 \,
   _2F_1\left(\frac{13}{6},\frac{17}{6};5;27x^2\right)\nonumber\\
&&-93760 \,
   _2F_1\left(\frac{7}{6},\frac{11}{6};4;27x^2\right)-39819780 x^4 \,
   _2F_1\left(\frac{19}{6},\frac{23}{6};6;27x^2\right)\Big)\,.
\eea
\end{widetext}
We expect that by using the contiguity relations valid for the hypergeometric functions these relations can be further simplified, even if unnecessary.

It is immediate to show the relations
\beq
\frac{G_i (x)+\delta_{i0}}{2x}=-I_r^{\pi (i)}(x)\,,\qquad i=0,\ldots 4\,,
\eeq 
which connect the null geodesics radial action (expanded in powers of $\hat a$) to the RAM ratio $\frac{\Delta \nu}{\nu}\sim \frac{\nu-l}{l}$ (also expanded in powers of $\hat a$).
Notice that the various $I_r^{\pi (i)}(x)$ have  specific parity properties ($I_r^{\pi (0)}(x)$ is $x$-odd, $I_r^{\pi (0)}(x)$ is $x$-even, etc.), and hence selecting the $\pi$-part of the radial action is equivalent to extract  specific parity functions from the full $I_r^{(i)}(x)$ (in the sense that this point of view is always available), as already discussed in Ref. \cite{Bini:2025ltr}.

\section{Concluding remarks}

Recent efforts  have been done in the literature \cite{Ivanov:2025ozg,Parnachev:2020zbr,Akpinar:2025huz,Bini:2025ltr} to study null particles radial action in the Kerr case as well as other relevant spacetimes and the \lq\lq renormalized angular momentum" appearing in the Mano-Suzuki-Takasugi formalism for the analytical treatment of gravitational perturbations following the Teukolsky approach.

We have computed the eikonal phase by using the MST solutions to the homogeneous radial (scalar) wave equation, providing a proof for the relation suggested by Ref. \cite{Ivanov:2025ozg} between the radial action for null geodesics and  the  renormalized angular momentum of the  MST  formalism.
We have also extended the linear-in-spin results of Ref. \cite{Ivanov:2025ozg} to higher orders in the Kerr rotational parameter $a$, finding explicit, nontrivial  resummation expressions (in terms of hypergeometric functions) up to the fourth order in $a$. These novel results are included in Tables \ref{tab:1} and \ref{tab:1bis} as well as in Eq. \eqref{variousG_i}.

Very likely, this paper is not the end of the story. In fact: 1) One should aim at using these resummed expressions in black hole perturbation theory, also improving so far the accuracy in (semi-)analytical simulations (for example, Effective-One-Body-based) of the two-bodies dynamics which are of fundamental importance in the modeling of  gravitational wave signal; 2) One should generalize the scalar wave (spin-$0$) results to the the generic spin $s$ case, as in the Teukolsky equation.

We will address these (and related) points in future works.

\section*{Acknowledgments}

We thank  M. Bianchi, T. Damour, P. Mastrolia  and J. Parra-Martinez for useful discussions at the beginning of th epresent project.  D.B. acknowledges
membership to the Italian Gruppo Nazionale per la Fisica
Matematica (GNFM) of the Istituto Nazionale di Alta
Matematica (INDAM).\\

\appendix

\section{MST recurrence: explicit solution in the eikonal limit}
\label{rec_rel}
We list below the explicit solution for the only nonvanishing coefficients of the MST recurrence relation as well as the RAM $\nu$ in the eikonal limit and truncating expressions at $O(x^4,q^2)$. Higher-order expressions are included in the Supplemental Material file.  
\begin{widetext}
\bea
a_{-4}&=&\Bigg[ \frac{l^5 (l^5-11 l^4+47 l^3-97 l^2+96 l-36)}{6 (-1+2 l)^2 (1+2 l) (2 l-3)^2 (2 l-5)}-\frac{i (l-3) (l-2) l^5 \left(l^2-3 l+1\right)}{3 (2 l-5) (2 l-3) (2 l-1)^2 (2
   l+1)}q\nonumber\\
   &-&\frac{(l-3) (l-2) l^5 \left(8 l^3-30 l^2+33 l-12\right)}{6 (2 l-5) (2 l-3)^2
   (2 l-1)^2 (2 l+1)}q^2 \Bigg]x^4, \nonumber\\
a_{-3}&=& \Bigg[-\frac{ i l^4 (l^3-5 l^2+8 l-4)}{3 (2 l-3) (1+2 l) (-1+2 l)^2}+\frac{-24 l^7+96 l^6-112 l^5+32 l^4}{24 (2 l-3) (2 l-1)^2 (2 l+1)}q+\frac{i \left(3 l^7-11 l^6+12 l^5-4 l^4\right)}{2 (2 l-3) (2 l-1)^2 (2 l+1)}q^2\Bigg] x^3\nonumber\\
&+&\Bigg[\frac{2 l^4 (3 l^3-12 l^2+14 l-4)}{3 (2 l-3) (1+2 l) (-1+2 l)^2} -\frac{4 i \left(l^7-3 l^6+2 l^5\right)}{(2 l-3) (2 l-1)^2 (2 l+1)}q+\frac{2 \left(-6 l^7+18 l^6-14 l^5+4 l^4\right)}{3 (2 l-3) (2 l-1)^2 (2 l+1)}q^2\Bigg]x^4,  \nonumber\\
a_{-2}  &=& \Bigg[-\frac{l^3 (l^2-2 l+1)}{ (1+2 l) (2 l-1)^2 }+\frac{i (l-1) l^3}{(2 l-1) (2 l+1)}q+\frac{(l-1) l^3}{(2 l-1) (2 l+1)}q^2\Bigg]x^2\nonumber\\
&+&\Bigg[
-\frac{2 i l^3 (l-1)}{ (4l^2-1) }-\frac{4 (l-1) l^4}{(2 l-1)^2 (2 l+1)}q+\frac{i (l-1) l^3}{(2 l-1) (2 l+1)}q^2\Bigg]x^3\nonumber\\
&+&\Bigg[-\frac{l^3 \left(32 l^{10}-80 l^9+704 l^8-2448 l^7+1170 l^6+4029 l^5-7499
   l^4+2072 l^3+118 l^2+462 l-180\right)}{3 (2 l-5) (2 l-1)^4 (2 l+1)^3 (2
   l+3)}\nonumber\\
   &+&\frac{i l^3 \left(16 l^9-32 l^8+1080 l^7-2532 l^6-2103 l^5+5013 l^4-2488
   l^3-802 l^2+48 l+180\right)}{3 (l-1) (2 l-5) (2 l-1)^2 (2 l+1)^3 (2 l+3)}q\nonumber\\
   &+&\frac{l^3 \left(16 l^9-16 l^8+1016 l^7-2308 l^6-2505 l^5+4579 l^4-1416 l^3-884
   l^2-282 l+180\right)}{3 (l-1) (2 l-5) (2 l-1)^2 (2 l+1)^3 (2 l+3)}q^2\Bigg]x^4, \nonumber\\ 
a_{-1}  &=& \Bigg[\frac{i l^2 }{ (1+2l)}+\frac{l^2}{2 l+1}q-\frac{i l^2}{2 (2 l+1)}q^2\Bigg]x
-\frac{2 l^2}{ (1+2 l) }x^2\nonumber\\
&+&\Bigg[\frac{i l^2 (8 l^8+4 l^7+238 l^6-133 l^5-580 l^4+243 l^3+39 l^2+66 l-36)}{ (1+2 l)^3 (-1+2 l)^2 (-9+4 l^2) }\nonumber\\
&+&\frac{l^2 \left(8 l^8+4 l^7+478 l^6-253 l^5-1176 l^4+537 l^3+173
   l^2-36\right)}{(2 l-3) (2 l-1)^2 (2 l+1)^3 (2 l+3)}q\nonumber\\
   &-&\frac{i l^2 \left(8 l^8+36 l^7+206 l^6-181 l^5-580 l^4+253 l^3+41 l^2+66
   l-36\right)}{2 (2 l-3) (2 l-1)^2 (2 l+1)^3 (2 l+3)}q^2\Bigg]x^3\nonumber\\
&+&\Bigg[-\frac{2 l^2 (8 l^8+4 l^7+478 l^6-253 l^5-1176 l^4+537 l^3+173 l^2-36)}{ (1+2 l)^3 (-1+2 l)^2 (-9+4 l^2) }\nonumber\\
&-&\frac{4 i \left(8 l^{12}-12 l^{11}+118 l^{10}-17 l^9-266 l^8+23 l^7-4 l^6+96
   l^5-36 l^4\right)}{(l-1) (l+1) (2 l-3) (2 l-1)^2 (2 l+1)^3 (2 l+3)}q\nonumber\\
   &+&\frac{2 \left(-16 l^{12}+40 l^{11}-412 l^{10}+234 l^9+844 l^8-557 l^7+89 l^6-77
   l^5+35 l^4\right)}{(l-1) (l+1) (2 l-3) (2 l-1)^2 (2 l+1)^3 (2 l+3)}q^2\Bigg]x^4,  \nonumber\\
a_0 &=& 1,  \nonumber\\
\eea
\bea
a_1 &=& \Bigg[\frac{l (l+1)i}{ (1+2 l)}-\frac{l^2}{2 l+1}q-\frac{i \left(l^2+l\right)}{2 (2 l+1)}q^2\Bigg] x
+\frac{2 l^2}{ (1+2 l)} x^2 \nonumber\\
&+&\Bigg[\frac{i l^3 (8 l^8+60 l^7+434 l^6+1925 l^5+4075 l^4+3835 l^3+870 l^2-908 l-511)}{ (1+2 l)^3 (3+2 l)^2 (-1+2 l) (l+1) (2 l+5)}\nonumber\\
&-&\frac{l^4 \left(8 l^8+60 l^7+674 l^6+3485 l^5+7679 l^4+7157 l^3+1346 l^2-1800
   l-841\right)}{(l+1)^2 (2 l-1) (2 l+1)^3 (2 l+3)^2 (2 l+5)}q\nonumber\\
   &-&\frac{i l^3 \left(8 l^8+60 l^7+434 l^6+1941 l^5+4115 l^4+3863 l^3+868 l^2-916
   l-513\right)}{2 (l+1) (2 l-1) (2 l+1)^3 (2 l+3)^2 (2 l+5)}q^2\Bigg] x^3\nonumber\\
&+&\Bigg[\frac{2 l^4 (8 l^8+60 l^7+674 l^6+3485 l^5+7679 l^4+7157 l^3+1346 l^2-1800 l-841)}{ (1+2 l)^3 (3+2 l)^2 (-1+2 l) (l+1)^2 (2 l+5)}\nonumber\\
&-&\frac{4 i l^4 \left(8 l^8+76 l^7+426 l^6+1425 l^5+2569 l^4+2311 l^3+747
   l^2-296 l-270\right)}{(l+1) (l+2) (2 l-1) (2 l+1)^3 (2 l+3)^2 (2 l+5)}q\nonumber\\
   &+&\frac{2 l^4 \left(16 l^9+152 l^8+1012 l^7+4042 l^6+8406 l^5+8444 l^4+2791
   l^3-1270 l^2-891 l+2\right)}{(l+1)^2 (l+2) (2 l-1) (2 l+1)^3 (2 l+3)^2 (2
   l+5)}q^2\Bigg] x^4,  \nonumber\\
a_2 &=& \Bigg[-\frac{l^2 (l^3+5 l^2+8 l+4)}{ (3+2 l)^2 (1+2 l)} -\frac{i l^3 (l+2)}{(2 l+1) (2 l+3)}q+\frac{l^2 (l+2) \left(2 l^2+3 l+2\right)}{(2 l+1) (2 l+3)^2}q^2\Bigg]x^2\nonumber\\
&+&\Bigg[\frac{2i l^3 (l+2)}{ (3+2l) (1+2l) }-\frac{4 \left(l^5+2 l^4\right)}{(2 l+1) (2 l+3)^2}q-\frac{i \left(l^4+2 l^3\right)}{(2 l+1) (2 l+3)}q^2\Bigg]x^3\nonumber\\
&+&\Bigg[-\frac{ l^4 (32 l^{10}+400 l^9+2864 l^8+14800 l^7+51458 l^6+111967 l^5+141666 l^4+86706 l^3+1608 l^2-25755 l-9690)}{ 3 (3+2 l)^4 (1+2 l)^3 (-1+2 l) (2 l+7) (l+1) }\nonumber\\
&-&\frac{i l^5 \left(16 l^9+176 l^8+1912 l^7+12332 l^6+39577 l^5+64508 l^4+48006
   l^3+4362 l^2-13227 l-5274\right)}{3 (l+1)^2 (l+2) (2 l-1) (2 l+1)^3 (2
   l+3)^2 (2 l+7)}q\nonumber\\
   &{+}&\frac{l^4 \left(32 l^{11}{+}400 l^{10}{+}4384 l^9{+}29696 l^8{+}109018 l^7{+}228071
   l^6{+}279713 l^5{+}191905 l^4{+}48441 l^3{-}29844 l^2{-}28668 l{-}7488\right)}{3 (l+1)^2
   (l+2) (2 l-1) (2 l+1)^3 (2 l+3)^3 (2 l+7)}q^2\Bigg]x^4,  \nonumber\\
a_3 &=& \Bigg[-\frac{i l^3 (l^4+9 l^3+29 l^2+39 l+18)}{3  (2 l+5) (3+2 l)^2 (1+2 l)} +\frac{24 l^7+168 l^6+376 l^5+264 l^4}{24 (2 l+1) (2 l+3)^2 (2 l+5)}q+\frac{i \left(3 l^7+19 l^6+41 l^5+39 l^4+18 l^3\right)}{2 (2 l+1) (2 l+3)^2 (2
   l+5)}q^2\Bigg]x^3\nonumber\\
&+&\Bigg[-\frac{2 l^4(3 l^3+21 l^2+47 l+33) }{3 (2 l+5) (3+2 l)^2 (1+2 l) }-\frac{4 i \left(l^7+5 l^6+6 l^5\right)}{(2 l+1) (2 l+3)^2 (2 l+5)}q+\frac{2 \left(6 l^7+30 l^6+47 l^5+33 l^4\right)}{3 (2 l+1) (2 l+3)^2 (2 l+5)}q^2\Bigg]x^4,  \nonumber\\
a_4 &=& \Bigg[\frac{ l^4 (l^6+17 l^5+117 l^4+415 l^3+794 l^2+768 l+288)}{ 6 (3+2 l)^2 (2 l+7) (2 l+5)^2 (1+2 l) }+\frac{i l^5 (l+3) (l+4) \left(l^2+5 l+5\right)}{3 (2 l+1) (2 l+3)^2 (2 l+5) (2
   l+7)}q\nonumber\\
   &-&\frac{l^4 (l+3) (l+4) \left(8 l^4+50 l^3+105 l^2+100 l+48\right)}{6 (2 l+1) (2
   l+3)^2 (2 l+5)^2 (2 l+7)}q^2\Bigg]x^4\,.
\eea

Moreover, the renormalized angular momentum is given by
\beq
\nu=l+ \sum_{k=2}^\infty \nu_k x^k\,.
\eeq
where
\bea
\nu_2 &=& -\frac{2 l^2 \left(15 l^2+15 l-11\right)}{(2 l-1) (2 l+1) (2 l+3)}\nonumber\\
\nu_3 &=& \frac{8 l^3 \left(5 l^2+5 l-3\right) q}{(l+1) (2 l-1) (2 l+1) (2 l+3)}\nonumber\\
\nu_4 &=& -\frac{4 \left(95 l^4+205 l^3-274 l^2-438 l+126\right) l^4 q^2}{(l+1) (2 l-3) (2 l-1)
   (2 l+1) (2 l+3)^2 (2 l+5)}\nonumber\\
&-&\frac{2  l^3}{(l+1)
   (2 l-3) (2 l-1)^3 (2 l+1)^3 (2 l+3)^3 (2 l+5)}(18480 l^{10}+92400 l^9+79800 l^8-235200
   l^7\nonumber\\
&-& 382305 l^6+64365 l^5+278260 l^4-9955 l^3-73892 l^2+8733 l+3240)\,. 
\eea

\end{widetext}

\end{document}